\documentclass{article}
\usepackage{spconf,amsmath,amssymb,amsfonts}
\usepackage{cite}
\usepackage{tabularray}
\usepackage{graphicx}

\usepackage{booktabs}
\usepackage{multirow}
\usepackage{subcaption}
\usepackage{xcolor}
\DeclareMathOperator*{\argmin}{arg\,min}

\title{An explainable proxy model for multilabel audio segmentation}
\name{Théo Mariotte$^1$ \thanks{This project has received funding from the European Union’s Horizon 2020 research and innovation program under the Marie Skłodowska-Curie grant agreement No 101007666. This work was performed using HPC resources from GENCI–IDRIS (Grant 2022-AD011012565). The research reported here was conducted at the 2023 Frederick Jelinek Memorial Summer Workshop on Speech and Language Technologies, hosted at Le Mans University (France) and sponsored by Johns Hopkins University.}, 
Antonio Almudévar$^2$, Marie Tahon$^1$, Alfonso Ortega$^2$}

\address{$^{1}$LIUM, Le Mans Université\\ $^2$ ViVoLab, Aragón Institute for Engineering Research (I3A), University of Zaragoza\\
theo.mariotte@univ-lemans.fr
}

\begin{document}
\ninept

\maketitle
\begin{abstract}
Audio signal segmentation is a key task for automatic audio indexing. 
It consists of detecting the boundaries of class-homogeneous segments in the signal.
In many applications, explainable AI is a vital process for transparency of decision-making with machine learning.
In this paper, we propose an explainable multilabel segmentation model that solves speech activity (SAD), music (MD), noise (ND), and overlapped speech detection (OSD) simultaneously.
This proxy uses the non-negative matrix factorization (NMF) to map the embedding used for the segmentation to the frequency domain.
Experiments conducted on two datasets show similar performances as the pre-trained black box model while showing strong explainability features.
Specifically, the frequency bins used for the decision can be easily identified at both the segment level (local explanations) and global level (class prototypes).
\end{abstract}

\section{Introduction}
Audio segmentation is a key task for many speech technologies such as automatic speech recognition, speaker identification, and dialog monitoring in different multi-speaker scenarios, including TV/radio, meetings, and medical conversations. 
More precisely, these technologies must be aware of the presence of noisy environments (brouhaha, external noise), and how many speakers are active at each time.
In many domains, such as health or human-machine interactions, the prediction of segment timestamps is not enough and it is necessary to include some explanations. 
Indeed, the current trend for explainable AI is a vital process for transparency of decision-making with machine learning: the user (a doctor, a judge, or a human scientist) has to justify the choice made based on the system output.
Explainability for AI can be addressed at different stages of the process. Pre-hoc explainability intends to understand and describe data with explainable features and statistics. 
Another stage is to develop explainable-by-design models. 
The last stage consists of the extraction of explanations from a pre-trained model by the use of proxy models or perturbation mechanisms. 
Our work comes within the scope of such post-hoc explanations.

The proposed approach aims to train an explainable proxy model from a black-box segmentation system.
While many works have been conducted in explaining neural networks in vision \cite{ribeiro_why_2016} and NLP \cite{murphy_learning_2012}, the literature is limited in the audio domain.
The first intents are focused on saliency map extraction to reveal what information is used in the output \cite{ribeiro_why_2016,qi2019visualizing}.
Audio classifier decisions can be interpreted by post-hoc analysis with approaches such as SoundLime \cite{mishra2017local}.
Recently, a few explainable models have been developed like APNet \cite{zinemanas2021interpretable}, which extends the training of prototypes to the audio domain, and post-hoc visualization of explanations obtained from Shapley values \cite{ge2022explaining}.
In the architecture proposed in \cite{parekh2023tackling}, the authors explain a black box audio classifier with a proxy model which is optimized to classify audio scenes while reconstructing the audio with the non-negative matrix factorization framework \cite{lee2000algorithms}.

%

This paper focuses on automatic audio segmentation in the context of large-scale multimedia data from archivists, i.e. the detection of homogeneous segments containing speech (SAD), music (MD), noise (ND), and overlapped speech (OSD) with a unified model.
%
Early studies on speech \cite{sohn1999statistical}, overlapped speech \cite{charlet_impact_2013}, and music \cite{lavner2009decision} segmentation are focused on the statistical modeling of handcrafted acoustic features.
Currently, the segmentation is mainly performed with neural networks and supervised learning.
%
While each task has been generally solved independently as a binary frame-wise classification task (SAD \cite{lavechin2019end}, OSD \cite{bullock_overlap-aware_2020,lebourdais22_interspeech}, MD \cite{jang2019music,de2019exploring}), more recent approaches propose to solve multiple tasks simultaneously. 
The multiclass model predicts a single class and class intersection is empty. 
For example, in \cite{gimeno2020multiclass}, authors propose to segment speech, music, and noise with a single multiclass model.
A few works also report joint SAD and OSD \cite{jung21_interspeech,bredin21_interspeech,lebourdais2023joint}.
In this paper, SAD, OSD, MD, and ND are simultaneously solved as a multilabel frame classification task.
Thus, multiple classes can be predicted simultaneously, and the intersection between classes is not empty.

We propose to train an explainable proxy model from a pre-trained multilabel segmentation model (designated as the teacher).
The architecture is inspired by \cite{parekh2023tackling}.
The proxy is trained following a teacher-student approach, commonly used in knowledge distillation \cite{hinton2015distilling}.
The teacher inputs Wavlm pre-trained features \cite{chen2022wavlm} and outputs the pseudo probability of each class.
Two types of proxy models are investigated.
The former inputs a spectrogram and the latter uses the teacher's Wavlm outputs.
Contrary to \cite{parekh2023tackling}, we consider frame-level, i.e. time segmentation, instead of utterance-level classification.
We demonstrate that proxy models provide similar or even better performance as the teacher while being smaller and explainable.
These models provide local (segment-level) explanations by highlighting the salient frequencies for a given audio input and global explanations that can be seen as class prototypes. 
 Contrary to state-of-the-art methods such as Shapley \cite{ge2022explaining} or SoundLime \cite{mishra2017local}, our contribution not only brings post-hoc explanations but also provides decisions.
Finally, we show that the relevant information for classification can be mapped to the spectral domain.
A classification score confirms the selection of relevant components.
To the best of our knowledge, this is the first intent to design an explainable neural multilabel segmentation model.  

This paper is arranged as follows: sec. \ref{sect:models} presents the system along with the proxy model training strategy.
Sec. \ref{sect:protocol} introduces the experimental protocol and sec. \ref{sect:perf} the segmentation results. 
Sec. \ref{sect:explain} demonstrates the explainability capacities of the proposed system.

\section{NMF-based explainable model}
\label{sect:models}

The architectures and training framework of the proxy models are described in this section.
\begin{figure}[ht]
    \centering
    \includegraphics[width=\linewidth]{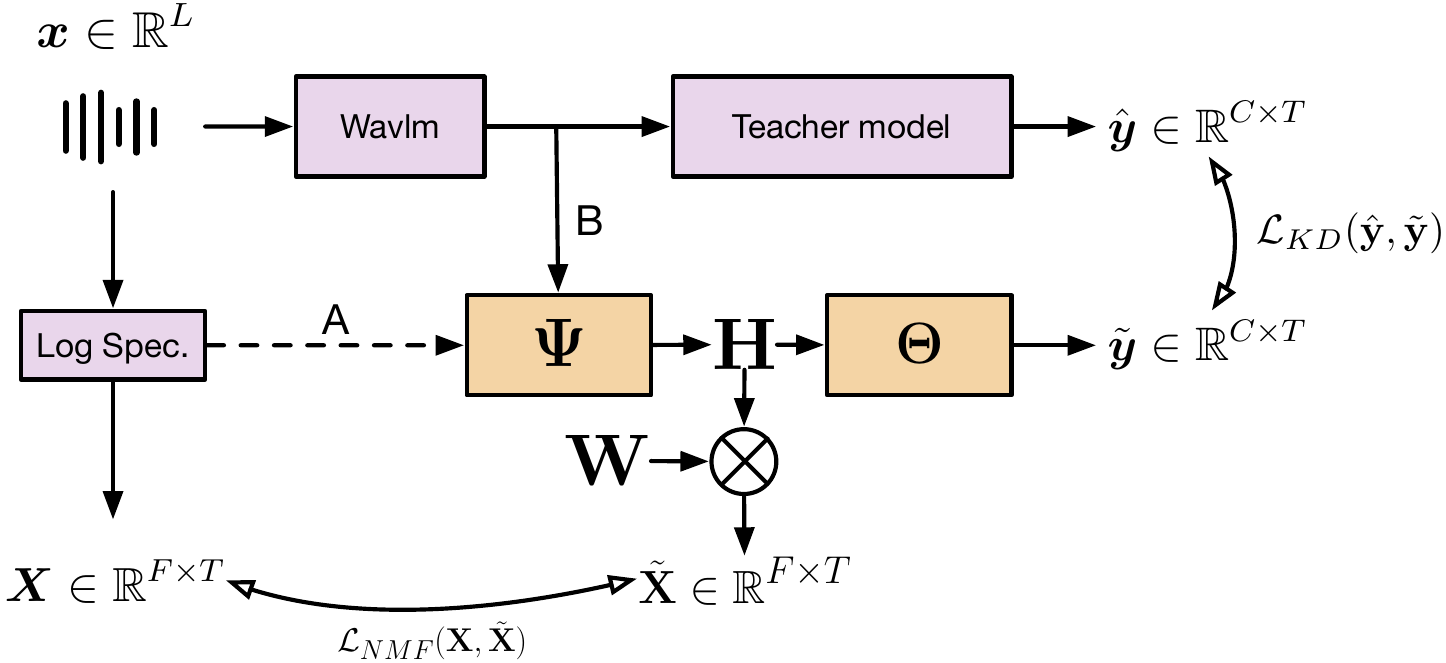}
    \caption{Diagram of the proposed architecture for NMF-based multi-label segmentation explanation. The \textbf{A} and \textbf{B} branches represent the spectrogram and Wavlm-based proxy models respectively.}
    \label{fig:enter-label}
\end{figure}
\vspace{-20pt}
\subsection{Multilabel audio segmentation formulation}

Audio segmentation is solved as a multilabel framewise classification task.
Let $\{\mathbf{S}, \mathbf{y}\}$ be a training set composed of sequences of acoustic features $\mathbf{S}\in\mathbb{R}^{D\times T}$, where $D$ is the feature vector dimension and $T$ the number of frames, and the aligned annotations $\mathbf{y}\in \mathbb{R}^{C\times T}$, with $C$ being the number of classes. 
Let $f:\mathbb{R}^{D\times T} \rightarrow \mathbb{R}^{C\times T}$ be a parametric function that estimates the logits for each class from the sequence of features such as $\hat{\mathbf{y}}=f(\mathbf{S})$.
The parameters of $f$ are optimized to minimize a loss function $\mathcal{L}(\hat{\mathbf{y}},\mathbf{y})$ (here we used a cross-entropy), with an iterative algorithm. The sequence of feature $\mathbf{S}$ is extracted from the audio signal by an additional function $\mathbf{S}=g(\mathbf{x})$. 

\subsection{Non-negative matrix factorization (NMF)}

NMF has been extensively used for audio signal processing \cite{richard2023audio}.
This approach factorizes a given non-negative matrix $\mathbf{X}\in\mathbb{R}_{+}^{F\times T}$ into two non-negative matrices: $\mathbf{W}\in\mathbb{R}_{+}^{F\times K}$, usually denoted as \textit{dictionary}, and $\mathbf{H}\in\mathbb{R}_{+}^{K\times T}$, denoted as \textit{activations}.
$K$ represents the rank of the factorization.
Both matrices are jointly learned by solving
\begin{equation}
    \mathbf{W},\mathbf{H} = \argmin_{\mathbf{W},\mathbf{H}} \|\mathbf{X}-\mathbf{W}\mathbf{H}\|_2^2,
\end{equation}
with a two-step optimization process \cite{lee2000algorithms}.
In this work, we consider the sparse NMF implementation \cite{le2015sparse} as suggested in \cite{parekh2023tackling}.
In our segmentation proxy model, the dictionary $\mathbf{W}$ is pre-learned while the activation $\mathbf{H}$ is extracted by a neural model $\Psi$ and referred to as an embedding.
The proxy model architecture and the way NMF is integrated is described in the following subsection.

\subsection{Proxy model framework}

In this work, the $f$ model is pre-trained with frozen weights and serves as a teacher for the proxy model.
We use a similar approach as \cite{parekh2023tackling} where the proxy model is composed of two functions.
Let $\Psi$ be a function that maps a sequence of $D$-dimension feature vectors $\mathbf{S}\in \mathbb{R}^{D\times T}$ to the embedding $\mathbf{H}\in \mathbb{R}_{+}^{K \times T}$. 
The proxy model logits $\tilde{\mathbf{y}}\in\mathbb{R}^{C\times T}$ are obtained with an additional $\Theta$ function such as $\tilde{\mathbf{y}}=\Theta(\mathbf{H})$.
The log-spectrogram of the input audio signal $\tilde{\mathbf{X}}$ is also reconstructed from $\mathbf{H}$ with the $\mathbf{W}$ dictionary: $\tilde{\mathbf{X}} = \mathbf{W}\mathbf{H}$.
Non-negativity is ensured by a ReLU activation function $\sigma$, e.g. $\sigma(\mathbf{H})$.
Note that $\mathbf{W}$ can also be trained simultaneously with the model.

The training objective of such a model is composed of 3 loss terms.
The first trains the proxy model to mimic the teacher's decisions.
Considering multilabel segmentation, and following the common approaches in knowledge distillation \cite{hinton2015distilling}, we use the binary Kullback-Leiber (KL) divergence between the teacher and the proxy model output distributions, denoted as $\mathcal{L}_{KD}(\hat{\mathbf{y}},\tilde{\mathbf{y}})$.

The second loss term constrains the $\mathbf{H}$ embedding to minimize the NMF-based spectrogram reconstruction and is implemented as the squared L2-norm between the target spectrogram $\mathbf{X}$ and the reconstruction $\tilde{\mathbf{X}}$:
\begin{equation}
    \mathcal{L}_{NMF}(\mathbf{X},\tilde{\mathbf{X}})=\|\mathbf{X}-\mathbf{W}\mathbf{H}\|_2^2.
\end{equation}

The last loss term minimizes the L1-norm of the $\mathbf{H}$ embedding to enforce its sparsity.
Having a sparse embedding reduces the number of active components, and makes the explanation step easier (Sec. \ref{sect:explain}).
Finally, the global training objective is the weighted sum of the 3 terms given in eq.~\ref{eq:full_loss} where $(\alpha, \beta, \gamma)$ is a triplet of hyperparameters to weight each term of the loss.
\begin{equation}\label{eq:full_loss}
    \mathcal{L} = \alpha \mathcal{L}_{KD}(\hat{\mathbf{y}},\tilde{\mathbf{y}}) + \beta \mathcal{L}_{NMF}(\mathbf{X},\tilde{\mathbf{X}}) + \gamma \|\mathbf{H}\|_1.
\end{equation}

\section{Experimental protocol}
\label{sect:protocol}
\subsection{Datasets}

In the current literature, there is no audio data annotated to perform SAD, OSD, MD, and ND simultaneously.
Therefore, models are trained on 4 datasets listed in the table \ref{tab:datasets} to perform multilabel audio segmentation so that all the classes are represented.
The teacher is optimized on the 4 training subsets described in the papers associated with each dataset.
This represents about 300 hours of annotated audio for training.
Due to the large amount of data, the teacher pre-training requires a lot of resources and a lengthy training process to converge.
We propose to train the proxy model on a subset of the training set.
Only Albayzin (AragonRadio train and 324TV) and DiHard train data are used for the knowledge distillation step, which represents 87 hours of audio.

To evaluate the segmentation performance, both teacher and proxy models are evaluated following the same protocol.
The models are evaluated on Aragon Radio and DiHard III test sets.
The test partitions follow the split proposed for the original challenges \cite{albayzin12,ryant2021dihard}.

\begin{table}[t]
    \centering
    \caption{Datasets used to train both teacher (T) and proxy (P) models with the available labels in each of them. Missing annotations for a given class are removed from the classification loss.}
    \begin{tabular}{l  cc  cccc}
         \toprule
         & \multicolumn{2}{c}{Model} & \multicolumn{4}{c}{Available label} \\
         \cmidrule(lr){2-3}
         \cmidrule(lr){4-7}
         Dataset & T & P & SAD & MD & ND & OSD \\
         \midrule
         Albayzin \cite{albayzin10,albayzin12} & \checkmark & \checkmark & \checkmark & \checkmark & \checkmark &  \\
         OpenBMAT \cite{openBMAT} & \checkmark & & & \checkmark & & \\
         ALLIES \cite{larcher:hal-03262914_short} & \checkmark & & \checkmark & & & \checkmark \\
         DiHard III \cite{ryant2021dihard} & \checkmark & \checkmark & \checkmark & & & \checkmark \\
         \bottomrule
    \end{tabular}
    \vspace{-10pt}
    \label{tab:datasets}
\end{table}

\subsection{Model architectures}

The teacher model is similar to \cite{lebourdais22_interspeech} and is composed of two main parts: feature extraction and sequence modeling.
The former is performed using pre-trained Wavlm Large \cite{chen2022wavlm} that outputs a sequence of 1024-dimension vectors.
The latter transforms the sequence of features to predict the segmentation.
It is composed of a 64-channel bottleneck layer followed by 3 TCN blocks \cite{bai_empirical_2018} composed of 5 1-D convolutional layers with exponentially increasing dilatation. 
A 1-D convolution layer processes the output of the TCN blocks to project the hidden representations to the logits space.
A sigmoid activation function is then applied to get normalized scores for each class.

As described in section \ref{sect:models}, two types of proxy models are investigated.
In the first approach, the proxy model (denoted as Spec. in Table~\ref{tab:seg_results}) inputs a log-spectrogram extracted on 64ms sliding windows with 20ms step.
In that case, $\mathbf{S}=\mathbf{X}$.
The $\Psi$ function is implemented with the same TCN-based architecture as the teacher but with 4 TCN blocks of 4 convolutional layers.
The bottleneck and hidden layers are composed of 128 and 256 channels respectively.
The second approach uses Wavlm features (denoted as Wavlm) from the teacher as input.
In this case, the $\Psi$ model is smaller to reduce the number of trainable parameters w.r.t. the teacher. 
The architecture is the same as the teacher's but with only 2 TCN blocks of 3 convolutional layers.
The $\Theta$ function is implemented as a linear layer with no bias of $K$ input dimensions and $C$ outputs, followed by a sigmoid.

\vspace{-5pt}

\subsection{NMF pre-training and hyperparameters}

As previously introduced, the $\mathbf{W}$ dictionary is pre-trained to map the embedding to the frequency domain.
The pre-training is performed on a subset of Aragon Radio using Sparse NMF \cite{le2015sparse}.
We select segments between 1s and 4s representing each type of class.
In practice, we found that building a subset with 1200 segments containing 16\% of speech and 42\% of music and noise offers the best reconstruction quality for each type of signal.
This approach however showed limited reconstruction quality with the Wavlm-based model.
To tackle this, we explored training the $\mathbf{W}$ dictionary along with the proxy model (denoted as Wavlm ($\mathbf{W}$ trained)).
In this case, the dictionary is implemented as a linear layer with no bias followed by a ReLU activation.
For each approach, the NMF rank is fixed at $K=256$ since it offers the best performance.

The proxy model is optimized with the ADAM optimizer with a learning rate of $10^{-3}$ and batches of 64 segments.
We set $(\alpha,\beta,\gamma)=(10,5,0.1)$ to scale the classification and reconstruction losses while lowering the impact of the sparsity term.


\section{Segmentation performance}
\label{sect:perf}

This section compares the performances in terms of the F1-score of each proxy model to the teacher model which is considered as our baseline.
The SAD, ND, MD, and OSD performances are presented in the table \ref*{tab:seg_results}.
The proxy models are expected to reach similar, eventually slightly lower, performances as the teacher.
The score obtained on OSD by the teacher model compares with the state of the art which is 63.4\% on DiHard III \cite{lebourdais22_interspeech}.

The spectrogram-based system (spec.) shows 96.3\% and 95.5\% F1-score on the SAD task on DiHard and AragonRadio respectively.
This represents a little degradation concerning the teacher.
The ND (73.0\%) and MD (87.6\%) are also degraded by an absolute -5,6\% compared to the teacher.
On OSD, the performance is strongly degraded with a 40.1\% F1-score.
The spectrogram-based model degrades the segmentation performance.
However, this result is expected since the teacher is trained on Wavlm pre-trained features which perform better than the spectrogram.
Considering the cost of Wavlm pre-training and the inference time, the spectrogram-based model offers a convincing performance.

\begin{table}[t]
    \centering
    \caption{F1-score (\%) obtained for each segmentation task on both AragonRadio and DiHard III evaluation sets. Bold values indicate the significantly best scores.}
    \begin{tabular}{l ccc cc}

        \toprule
         & \multicolumn{3}{c}{Aragon Radio} & \multicolumn{2}{c}{DiHard III} \\
         \cmidrule(lr){2-4}
         \cmidrule(lr){5-6}
         Model & SAD & ND & MD & SAD & OSD \\  
         \midrule
         Teacher & \textbf{96.8} & 78.6 & \textbf{93.2} & \textbf{96.9} & 60.7\\
         \midrule
         Spec. & 96.3 & 73.0 &  87.6 & 95.5 & 40.1\\
         Wavlm ($\mathbf{W}$ trained) & 96.7 & 78.5 & 93.0 & 96.8 & 60.8\\
         Wavlm & \textbf{96.8} & \textbf{79.5} & \textbf{93.1} & \textbf{96.9} & \textbf{61.4} \\
         \bottomrule
    \end{tabular}
    \vspace{-15pt}
    \label{tab:seg_results}
\end{table}

Using Wavlm features as input of the proxy model drastically improves the segmentation performance.
When the NMF dictionary is trained simultaneously as the segmentation task, the proxy model reaches 96.1\% and 96.7\% F1-score on SAD.
The ND performance remains lower than the teachers with a 72.6\% F1-score, which represents an absolute -6\% degradation.
The MD and OSD performances are similar to the teacher with 93.0\% and 60.5\% F1-score respectively.
While showing similar performance as the teacher, training the NMF dictionary degrades noise detection.
When the NMF dictionary is pre-learned, the proxy model delivers similar or even better segmentation performance as the teacher.
The proxy model reaches a 79.5\% F1-score on ND (+0.9\% absolute improvement w.r.t. the teacher), and 61.4\% F1-score on OSD.
This system shows the best segmentation performance among all the models.

The Wavlm-based proxy models deliver the best performance when the dictionary is pre-learned.
This model is kept for explanation extraction in the next section.


\section{Decision explanation}
\label{sect:explain}
This section describes the explanation extraction process to identify the relevant frequency bins for the segmentation.

\subsection{Explanation extraction}

The NMF framework allows to map the $\mathbf{H}=[\mathbf{h}_1,\cdots,\mathbf{h}_t,\cdots,\mathbf{h}_T]$ where $\mathbf{h}_t \in \mathbb{R}^K$ embedding to the frequency domain.
Furthermore, the segmentation prediction is obtained from this embedding with the $\Theta$ linear transformation.
The first step in the explanation process is to identify the $k\in [1,\cdots, K]$ NMF components that are the most relevant for each classification task.
We first apply a pooling operation to the embedding by averaging it over the time dimension: $z_k = \frac{1}{T}\sum_{t=1}^T h_{k,t}$.
To identify the relevant components to detect the class $c$, we define a relevance vector $\mathbf{r}_c=[r_{1,c}, \cdots, r_{k,c}, \cdots r_{K,c}]$ in which each element is computed following \eqref{eq:relevance}, where $\theta_{k,c}$ is the $k$-th weight of the linear layer associated to class $c$.
The highest values in $\mathbf{r}$ correspond to the most relevant component.
Therefore, the most relevant components are selected by applying a threshold $\tau$. 
A filtered relevance vector $\mathbf{R}_{c,\tau}$ is obtained, in which the $k$-th element is defined as:
\vspace{-5pt}
\begin{equation}\label{eq:relevance}
    R_{k,c}(\tau) = 
    \begin{cases}
        r_{k,c} = z_k \times \theta_{k,c} & \text{if } r_{k,c} > \tau \\
        0       & \text{otherwise}. \\
    \end{cases}
\end{equation}

\subsection{Segment-level (local) explanation}

The filtered relevance vector $\mathbf{R}_{c}(\tau)$ belongs to the same space as the $\mathbf{H}$ embedding.
Hence, it can be projected to the frequency domain by the following NMF linear transformation:
\vspace{-5pt}
\begin{equation}
    \mathbf{X}_{c}(\tau) = \mathbf{W}\mathbf{R}_{c}(\tau),
    \label{eq:rel_proj_freq}
\end{equation}

where $\mathbf{X}_{c}(\tau)\in\mathbb{R}^{K}$ is the projection of the relevant components in the frequency domain given class $c$ and threshold $\tau$.
This representation highlights the relevant frequency bins to detect class $c$.

After identifying the relevant components, it is necessary to measure how confident the model is regarding this selection.
Model confidence is evaluated with the classification scores.
The score is obtained by only keeping the relevant components in the $\mathbf{H}$ embedding and forwarding this filtered embedding $\mathbf{H}_{c}(\tau)$ through the $\Theta$ layer such that $\tilde{\mathbf{y}}_{c}(\tau)=\Theta\left(\mathbf{H}_{c}(\tau)\right)$.
Thus, the frequency-domain explanation $\mathbf{X}_{c}(\tau)$ can be compared to the model output for a given threshold $\tau$.

\begin{figure}[t]
    \centering
    \includegraphics[width=\linewidth]{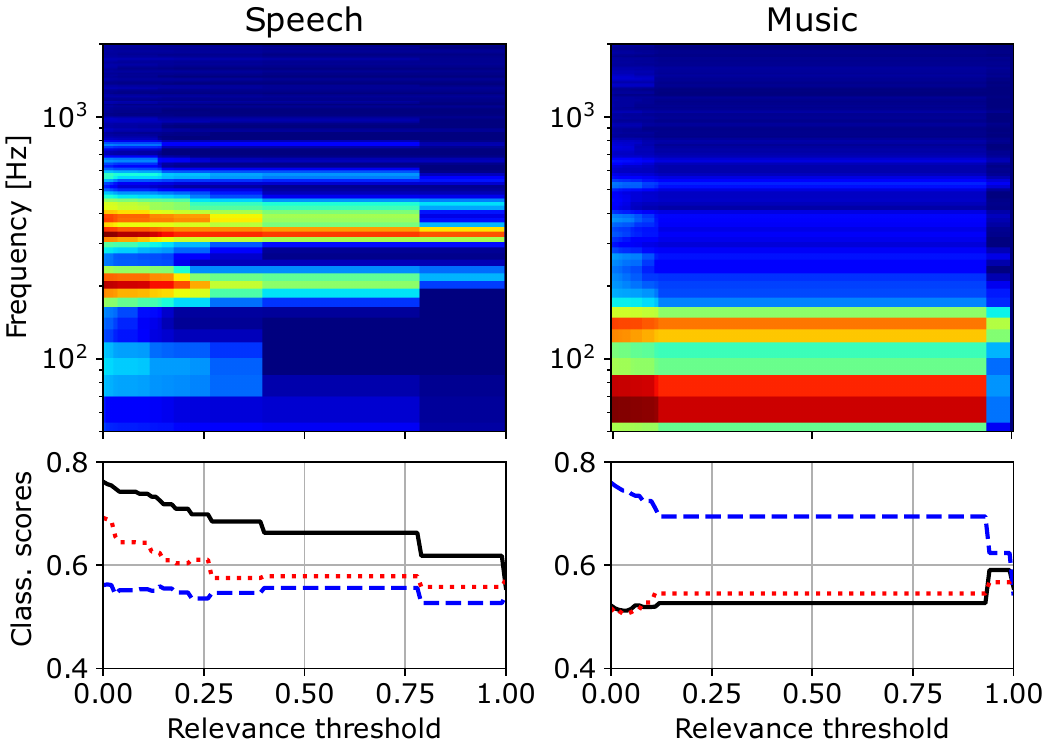}
    \caption{(Top) segment-level (local) relevant frequency bins for speech and music segmentation according to the relevance threshold $\tau$. (Bottom) Classification scores for (\textbf{--}) speech, (\textbf{\textcolor{blue}{-~-}}) music, and (\textcolor{red}{$\boldsymbol{\cdots}$}) overlapped speech with each selected components. Two audio samples from AragonRadio eval set with speech only (left) and music only (right)}
    \label{fig:seg_explain}
    \vspace{-15pt}
\end{figure}

Figure \ref{fig:seg_explain} shows the relevant components projected in the frequency domain with \eqref{eq:rel_proj_freq} for two types of segments: speech and music.
The relevant components are presented as a function of the relevance threshold: the higher $\tau$ the fewer components are selected. 
The classification score obtained from the selected components is also presented.
The figure shows that the relevant components for speech are located between 100Hz and 1kHz. 
While $\tau > 0.40$, the most relevant components for speech classification remain, \textit{e.g.}, around 200Hz.
As we are in a multilabel framework, the presence of speech can also imply the presence of overlap, therefore overlap prediction is also affected by these components.
This can be explained by the presence of common relevant frequency bins between SAD and OSD (Fig. \ref{fig:global_exp}) as described in the next section.

In the case of MD, the relevant frequency bins are located between 50Hz and 200Hz.
The components located in the band [60,80]Hz are highly relevant for music detection.
When those are removed ($\tau$=0.9), the music prediction drops.
Speech and overlap scores remain similar for each $\tau$ since the frequency bins are different between music and these classes.

The quality of the reconstruction in the frequency domain limits the explanation extraction.
In the case of the Wavlm-based proxy model, the reconstruction is of low quality in high frequencies. 
Thus, no explanation can be extracted in this part of the spectrum.

\subsection{Global explanation}

The previous explanation is obtained at the segment level, meaning that the relevant components are identified locally.
However, there is no guarantee that the components used to detect a class $c$ in a segment are the same across an entire dataset.
Thus, we propose a global explanation, $\bar{\mathbf{r}}_{c}$ vector, which highlights the global relevant components to detect the class $c$. It averages the relevance vectors $\mathbf{r}_{c}$ for a set of $N$ segments $\mathcal{D}_c=\{\mathbf{x}_{1,c}, \cdots, \mathbf{x}_{n,c}, \cdots \mathbf{x}_{N,c}\}$ containing the class of interest $c$: $\bar{\mathbf{r}}_{c} = \frac{1}{N}\sum_{\mathbf{x}_{c}\in\mathcal{D}_c} \mathbf{r}_c$.

\begin{figure}[t]
    \centering
    \includegraphics[width=0.8\linewidth]{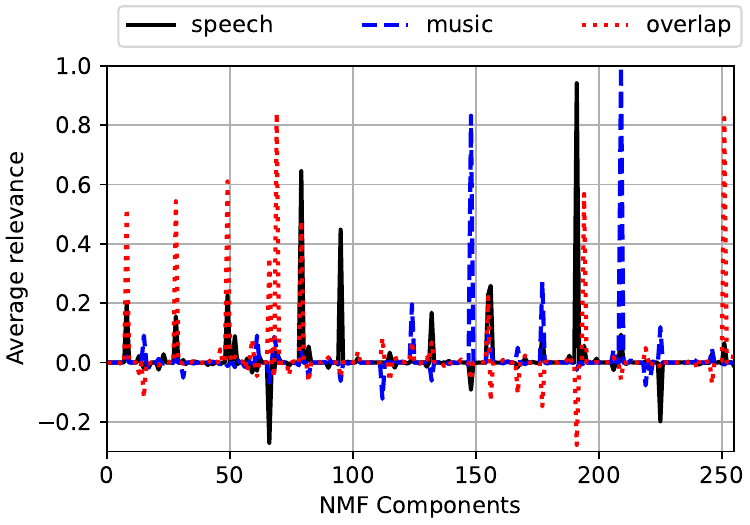}
    \caption{Global relevant components for speech (sp), music (mu), and overlapped speech (ov).}
    \label{fig:global_exp}
    \vspace{-15pt}
\end{figure}

To compute the average relevance $\bar{\mathbf{r}}_{c}$, $N=100$ segments containing exclusively each class of interest are selected to build each subset $\mathcal{D}_c$.
To better balance the data towards the overlap class, two-speaker artificial speech mixtures are generated by summing two randomly selected speech segments.
The noise class is not represented since too few segments contain exclusively noise.

Figure \ref{fig:global_exp} presents the average relevance $\bar{\mathbf{r}}_c$ for each $k$ component,  computed for SAD, OSD, and MD.
It shows that some components are typical for SAD and OSD.
For example, the component of index 50 is activated for both classes.
The figure also demonstrates that the components related to music are different from speech and overlap.
This confirms the behavior observed at the segment level in Fig. \ref{fig:seg_explain}.
Finally, Fig. \ref{fig:global_exp} shows negative values that are due to negative weights in the $\Theta$ layer.
This may highlight some components that are very discriminant between classes, e.g. around $k=70$ where the relevance is negative for speech and positive for overlap. 

\section{Conclusions}
\label{sect:ccl}
This paper proposes a set of explainable proxy models for multilabel audio segmentation.
The proxy is trained to fit the logit distribution of a pre-trained black-box teacher model.
Non-negative Matrix Factorization (NMF) is the core of this approach.
It maps the embedding used for the decision-making process to the frequency domain.
Hence, one can easily identify the relevant frequency bins used to predict the segmentation.
Experiments conducted on AragonRadio and DiHard III datasets show that the proxy model offers the same performance as the teacher while providing strong explainability capacities.
In the future, we plan to work on the evaluation of the extracted explanations

\newpage
\bibliographystyle{IEEEbib}
\bibliography{habi}
\end{document}